\documentclass[runningheads]{llncs}
\usepackage[T1]{fontenc}
\usepackage{booktabs}
\usepackage{graphicx}
\usepackage{multirow} 
\usepackage{array}
\usepackage{tabularx}
\usepackage{xcolor}

\begin{document}
\title{Why Machines Misread Pedagogical Quality: Human–Machine Alignment in LLM-Based Pretest Question Evaluation}
%
%
%
\author{Pei-Yu Tseng\inst{1}\orcidID{0000-0001-9675-674X} \and Mahir Akgun\inst{1}\orcidID{0000-0001-6884-0119} \and Peng Liu\inst{1}\orcidID{0000-0002-5091-8464}}

\authorrunning{P.-Y. Tseng et al.}

\institute{1 College of Information Sciences and Technology, The Pennsylvania State University, University Park, PA 16802, USA\\
\email{jerry950909@gmail.com, \{mza149,pxl20\}@psu.edu}}

\maketitle              
\begin{abstract}
Designing effective pretest questions is challenging at scale: high-quality questions require careful calibration of openness, cognitive depth, and alignment with learning objectives, yet generating and evaluating them manually is time-consuming. We present an AI-assisted workflow for pretest question development that combines automated generation, rubric-based evaluation, and iterative selection. Because the workflow relies on machine evaluation to filter questions at scale, we investigate the alignment between human and machine judgments across a 2×2 design varying rubric operationalization and evaluation mode. Our findings show that human-machine disagreements are systematic rather than random, that rubric revision has a larger effect on alignment than rationale-first evaluation, and that the two interventions are complementary. These findings highlight that scalable AI-assisted pretesting depends not only on generation capability but on how pedagogical quality is operationalized for machine interpretation.

\end{abstract}

\section{Introduction}
Pretesting—asking learners to answer questions before formal instruction—has emerged as a powerful strategy for improving learning. Although learners often generate incorrect responses during pretests, attempting to answer questions prior to studying can prime attention, activate relevant knowledge, and enhance encoding of subsequently presented information \cite{carpenter2018prequestions,pan2021pretesting}.  A growing body of research shows that such “errorful generation” can improve retention and sometimes transfer across a range of learning materials, including texts, lectures, and videos \cite{pan2021pretesting,richland2009pretesting}. However, designing effective pretest questions remains a significant challenge for instructors. High-quality questions are characterized by appropriate openness — ranging from constrained to fully open-ended response formats \cite{rodriguez2003construct,bennett1990toward} — cognitive depth (i.e., complexity) \cite{webb1997criteria}, and alignment with learning objectives \cite{webb1997criteria}, yet generating such questions at scale is time-consuming and requires pedagogical expertise \cite{kurdi2020systematic}.

Recent advances in generative AI offer new opportunities to support the creation of instructional materials, including pretest questions. Large language models can rapidly produce candidate questions, but their outputs vary widely in pedagogical quality. As a result, a key challenge is not only generating questions but also evaluating and selecting those that best support learning.

In this paper, we present an AI-assisted workflow for pretest question development consisting of three components: (1) automated generation grounded in instructional materials, (2) rubric-based evaluation of question quality, and (3) iterative selection of high-quality questions. Because the workflow relies on automated evaluation to filter questions at scale, the validity of that evaluation is critical. We therefore investigate the alignment between human and machine judgments in rubric-based pretest question evaluation, examining what drives disagreement and how it can be reduced. Using a 2×2 design, we vary rubric operationalization — comparing an initial human-established rubric against a revised rubric designed to make pedagogical constructs more explicit for machine use — and evaluation mode — comparing direct scoring against rationale-first evaluation, where the machine first articulates its reasoning for each dimension before assigning a score. Our findings shed light on when machine evaluators can reliably assess pedagogical quality, with implications for rubric design and evaluation prompting in AI-assisted assessment.


\section{Question Generation Workflow}
\label{Question Generation Workflow}

\subsection{Inputs of workflow}
\vspace{-6.5mm}
\begin{table*}[h!]
\centering
\renewcommand{\arraystretch}{0.85}
\setlength{\extrarowheight}{0pt}
\caption{Human-established rubric used in the initial experiment.}
\label{tab:initial_rubric}
\resizebox{\textwidth}{!}{%
\scriptsize
\begin{tabularx}{\textwidth}{p{0.8cm} X X X}
\toprule
\textbf{Dim.} & \textbf{Level 1} & \textbf{Level 2} & \textbf{Level 3} \\
\midrule
Open.
& One or two provided options (e.g., True/False). 
& Three or more provided options 
& Fully open-ended\\
\addlinespace[2pt]
Depth 
& Recall, recognition, definition, or basic interpretation without applying concepts to new contexts. 
& Application; interpreting statistics, model behavior, or predictor properties in a specific scenario. 
& Evaluation, comparison, justification, or strategic decision-making; requires integrating multiple concepts or reasoning about trade-offs. \\
\addlinespace[2pt]
Relev.
& Weakly related to the learning goals; off-topic or tangential. 
& Somewhat related; supports context but is not central to the main learning outcomes. 
& Strongly aligned with key lesson objectives such as feature selection, predictor evaluation, model performance, and regularization. \\
\addlinespace[2pt]
Clar.
& Unclear or misleading; requires guessing meanings, inferring missing information, or untangling multiple tasks. 
& Answerable but not clean; wording is unnecessarily long, vague, or assumes unstated background knowledge. 
& Immediately understandable; key terms are standard or explained, task expectations are explicit, and no clarification is needed. \\
\bottomrule
\end{tabularx}%
}
\end{table*}
\vspace{-5mm}
The workflow requires four inputs. The first three define the content scope of the generated questions: course material provides the conceptual content, a scenario situates the questions in a professionally relevant context, and key concepts constrain the topical scope. The fourth input is a rubric that defines the quality dimensions along which candidate questions are generated and evaluated: openness, cognitive depth, relevance, and clarity (see Table~\ref{tab:initial_rubric}). Relevance and clarity are well-established dimensions in question quality evaluation~\cite{gorgun2024exploring,kurdi2020systematic}. By contrast, openness and depth require more careful operationalization, as their intended meaning in pretest contexts can be misread by both human and machine raters. We define openness along a three-level scale based on the breadth of the response space: from selection among a fixed set of options, through constrained but non-fixed responses, to fully open-ended answers where multiple responses can be equally valid~\cite{bennett1990toward,rodriguez2003construct}. Depth follows Bloom's Taxonomy~\cite{krathwohl2002revision} across three levels: (1) recall or understanding, (2) application, and (3) analysis in context or evaluation. The full rubric, including relevance and clarity, is shown in Table~\ref{tab:initial_rubric}.

\subsection{System Overview}
\vspace{-7mm}
\begin{figure}[h!]
\centering
\includegraphics[width=\linewidth]{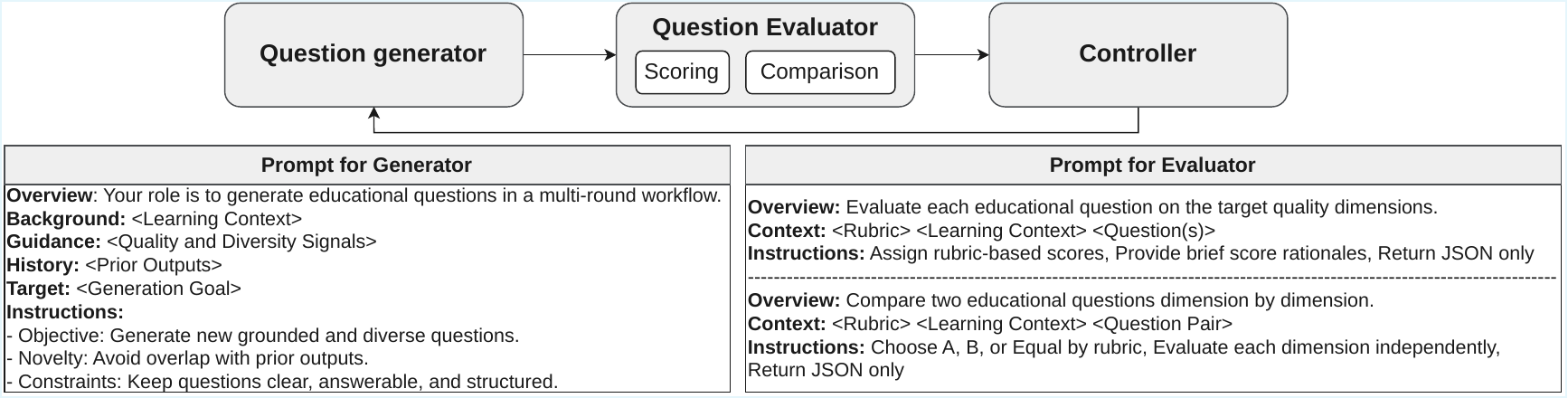}
\caption{Overview of the question generation workflow with prompt templates.}
\label{overview_and_prompt}
\end{figure}
\vspace{-5mm}

The workflow consists of three components that operate iteratively (Figure~\ref{overview_and_prompt}). In each round, the \textbf{Generator} receives all four inputs and, after the first round, feedback from the \textbf{Controller}, and outputs a batch of candidate questions. The \textbf{Evaluator} first scores each question individually on all rubric dimensions, removing those below level~3 on relevance or clarity. It then pairs the remaining questions for blind pairwise comparison against the same rubric; if a pairwise judgment contradicts the score-based ranking on any dimension, both questions are flagged as unreliable and removed. The \textbf{Controller} then analyzes the filtered questions to identify gaps in quality and topic coverage, including which question types and openness--depth cells have been overproduced or remain underrepresented, and passes this feedback to the \textbf{Generator} for the next round. This multi-round design, rather than a single-batch process, allows the system to improve diversity and coverage by filling gaps in the $3\times3$ openness--depth space while reducing redundancy across rounds.

\section{Rubric Design and Evaluator Reasoning in AI-Assisted Pretest Question Development}

\subsection{Overview and Experimental Setup}
The study is situated in an intermediate-level statistics course in the Cybersecurity major, and the generated pretest questions focus on multiple regression for undergraduate learners in that program. To evaluate the workflow's automated evaluation component, we use two factors in a $2 \times 2$ design. The first factor is \textit{rubric version}, comparing an initial rubric designed by human instructors (see Table~\ref{tab:initial_rubric}) with a revised rubric reconstructed to make pedagogical constructs more explicit for machine use. Since the rubric guides both generation and evaluation (see Section~\ref{Question Generation Workflow}), each rubric version produces its own set of candidate questions. The second factor is \textit{evaluation mode}, comparing direct evaluation, where the machine assigns scores immediately, against rationale-first evaluation, where the machine must first articulate reasoning for each dimension before assigning a score~\cite{chiang2023closer,liu2023geval}.

Both the generator and evaluator use GPT-5.1. This generation and filtering process was run separately under each rubric version, since the rubric guides generation as well as evaluation. In each run, the generator produces 12 candidate questions per round across 15 rounds, yielding 180 candidates. After the filtering process described in Section~\ref{Question Generation Workflow}, we retained approximately 120 questions per rubric version, selecting for diversity and aiming to keep coverage across the $3\times3$ openness--depth space as balanced as possible. Each question pool was then scored by the machine evaluator under both evaluation modes---direct and rationale-first---giving four conditions in total. To obtain human reference ratings, two independent raters, both experienced instructors in teaching applied statistics courses, scored a sample of 60 randomly selected questions under each rubric version. Like the machine evaluator, the human raters were given the same four workflow inputs---the rubric, scenario, learning material, and key concepts---together with the generated question, and assigned one level for each rubric dimension. Inter-rater agreement was substantial under the initial rubric (Cohen's $\kappa$ = 0.73) and improved under the revised rubric (Cohen's $\kappa$ = 0.78), suggesting that the rubric revision made the constructs more consistently interpretable for human raters as well. Following each inter-rater check, one rater scored the remaining questions under the corresponding rubric version. Human ratings are compared against machine ratings across all four conditions in terms of agreement in these per-dimension level judgments, using four metrics: Bias (Human $-$ Machine, capturing direction), MAE (average magnitude), MSE (penalizing larger errors), and Agree\% (proportion of exact matches). Results are reported in Table~\ref{tab:exp_aggregate}.

We begin with results under the initial rubric with direct evaluation (Exp.~1), then analyze why discrepancies arise and distill design principles for rubric revision, before examining the effects of the revised rubric (Exp.~2) and rationale-first evaluation (Exp.~3 and 4).

\subsection{Experiment~1: Initial Rubric, Direct Evaluation}
As shown in Table~\ref{tab:exp_aggregate} (Exp.~1), openness showed the largest discrepancy: its positive bias indicates systematic underestimation by the machine, and its low agreement (55.7\%) suggests it was the most difficult dimension for machine rating. Depth showed no directional bias but moderate agreement (70.5\%). By contrast, relevance and clarity showed near-perfect alignment, as both involve more surface-tractable judgments such as semantic matching and language quality that are more straightforward to operationalize in a rubric.

\vspace{-5mm}
\begin{table*}[h!]
\centering
\caption{Human--machine discrepancies across all four conditions. Init.\ = Initial rubric; Rev.\ = Revised rubric; Dir.\ = Direct; Rat.\ = Rationale-first.}
\resizebox{\textwidth}{!}{%
\setlength{\tabcolsep}{3pt}
\begin{tabular}{l|rrrr|rrrr|rrrr|rrrr}
\toprule
& \multicolumn{4}{c|}{\textit{Exp.~1: Init., Dir.}}
& \multicolumn{4}{c|}{\textit{Exp.~2: Rev., Dir.}}
& \multicolumn{4}{c|}{\textit{Exp.~3: Init., Rat.}}
& \multicolumn{4}{c}{\textit{Exp.~4: Rev., Rat.}} \\
\textbf{Dim.}
& \textbf{Bias} & \textbf{MAE} & \textbf{MSE} & \textbf{Ag.\%}
& \textbf{Bias} & \textbf{MAE} & \textbf{MSE} & \textbf{Ag.\%}
& \textbf{Bias} & \textbf{MAE} & \textbf{MSE} & \textbf{Ag.\%}
& \textbf{Bias} & \textbf{MAE} & \textbf{MSE} & \textbf{Ag.\%} \\
\midrule
Open.  & 0.66 & 0.66 & 1.08 & \textbf{55.7}  & 0.22 & 0.22 & 0.32 & \textbf{83.3}  & 0.37 & 0.37 & 0.37 & \textbf{63.0}  & 0.05 & 0.05 & 0.05 & \textbf{94.5} \\
Depth  & 0.00 & 0.30 & 0.30 & \textbf{70.5}  &-0.02 & 0.05 & 0.05 & \textbf{95.0}  &-0.06 & 0.13 & 0.13 & \textbf{87.0}  &-0.05 & 0.05 & 0.05 & \textbf{95.0} \\
Relev. & 0.00 & 0.00 & 0.00 & 100.0           & 0.00 & 0.00 & 0.00 & 100.0           & 0.00 & 0.00 & 0.00 & 100.0           & 0.00 & 0.00 & 0.00 & 100.0 \\
Clar.  &-0.02 & 0.02 & 0.02 & 98.4            & 0.00 & 0.00 & 0.00 & 100.0           &-0.02 & 0.02 & 0.02 & 98.1            & 0.00 & 0.00 & 0.00 & 100.0 \\
\bottomrule
\end{tabular}%
}
\label{tab:exp_aggregate}
\end{table*}
\vspace{-10mm}
\subsection{Mismatch Analysis, Rubric Revision, and Experiment~2}
The discrepancies in Experiment~1 were not random: they reflected recurring differences between the machine's scoring heuristics and the intended meaning of the rubric criteria.

\noindent\textbf{Openness.}
The machine tended to underestimate openness when a question appeared to have a standard answer, such as concept-definition, implication, or true/false questions with justification. It treated these as low-openness because they pointed toward a single answer, but the rubric was meant to capture whether students had to generate their own wording or justification, rather than select among fixed options. To address this, the revised rubric applied two principles. (P1)~It \textit{separated the intended construct from misleading surface cues}, making clear that openness depends on response-space breadth rather than on whether a question appears to have a single correct answer. (P2)~It \textit{specified what counts as valid evidence for each score level}, clarifying whether students remain within a bounded option space or are required to construct their own reasoning.
 
\noindent\textbf{Depth.} For depth, we identified three recurring patterns. (i)~The machine treated explanatory wording such as ``explain'' or ``why'' as evidence of analytical reasoning, even when the question only required conceptual understanding. (ii)~It treated procedural articulation as evaluative synthesis, over-scoring questions that only asked students to outline an approach. (iii)~It also underestimated short or binary questions that still required evaluating assumptions or judging validity. These patterns motivated two revision principles. (P3)~The revised rubric \textit{made the scoring logic operational rather than purely descriptive}: each level was defined through a classification structure with key characteristics, examples, and an explicit decision rule, rather than broad cognitive terms. (P4)~It \textit{made adjacent score boundaries explicit}, distinguishing interpretation without decision from decision with justification.

\noindent\textbf{Outcome: Revised Rubric (Exp. 2).} As shown in Table~\ref{tab:exp_aggregate}, openness agreement increased from 55.7\% to 83.3\%, bias decreased from 0.66 to 0.22, and MSE from 1.08 to 0.32. For depth, agreement increased from 70.5\% to 95.0\%, with very small remaining error. These gains confirm that the original mismatch was largely attributable to rubric operationalization rather than model inconsistency.


\subsection{Experiments~3 and~4: Effect of Rationale-First Evaluation}

We next tested whether rationale-first evaluation further improves alignment.
Rationale-first evaluation did not simply ask the model to “explain its score.” Instead, it inserted an intermediate rubric-grounded analysis step before final level assignment. For openness, the evaluator was required to enumerate distinct plausible answer directions, estimate the breadth of the answer space, and reflect on whether these directions were substantively distinct or only surface variants. For depth, the evaluator was required to pass through a decision gate and classify the question’s cognitive demand before mapping that analysis to a final level. This design was intended to reduce reliance on surface cues (e.g., the presence of “why” or “explain”) and force the evaluator to ground its judgments in rubric-relevant evidence.

\noindent\textbf{Under the initial rubric (Exp. 3).}
Rationale-first evaluation improved depth agreement from 70.5\% to 87.0\%, but openness agreement increased only from 55.7\% to 63.0\%, remaining substantially below the 83.3\% achieved by rubric revision alone (Experiment~2).
 
\noindent\textbf{Under the revised rubric (Exp. 4).}
Combining rationale-first evaluation with the revised rubric yielded the highest alignment overall. Openness agreement reached 94.5\% (MAE and MSE both 0.05). Depth agreement remained at 95.0\%, with no additional gain from rationale-first evaluation.
 
\noindent\textbf{Interpretation.} These results show that explicit reasoning alone cannot compensate for ambiguous construct definitions. Under the initial rubric, rationale-first evaluation substantially improved depth but yielded only limited gains in openness, indicating that prompting the model to “think more” is insufficient for poorly specified constructs. Once the rubric was operationalized, however, rationale-first evaluation produced substantial additional gains for openness, while offering little further benefit for depth, which had already reached high alignment. This asymmetry suggests that the two constructs impose different demands on the evaluator. Depth benefits from explicit decision structures—introduced through rubric design or rationale-first evaluation—though alignment is most stable when these structures are embedded in the rubric itself. In contrast, openness requires both a clear definition of the response space and active analysis of that space for a given question. Accordingly, it benefits most from combining a well-specified rubric with rationale-first evaluation, which supports exploration of multiple plausible answer paths.

\section{Conclusion}
Our findings suggest that, for LLM-based evaluation, providing operationalized criteria with explicit decision boundaries plays a more critical role than prompting the model to reason more carefully. At the same time, the two interventions are complementary, with the highest human–machine alignment achieved when combining an operationalized rubric with rationale-first evaluation.
This study focuses on alignment between human and machine judgments in rubric-based pretest question evaluation, rather than downstream educational impact. We therefore do not claim that improved alignment necessarily yields better pretest questions, better teaching decisions, or improved student learning outcomes. In addition, we compare two evaluation modes within a single LLM-based workflow rather than benchmarking a broader set of alternative evaluation methods. Future work should examine downstream learning effects and compare this workflow against a wider range of automated and hybrid evaluation baselines.
\bibliographystyle{splncs04}
\bibliography{reference}
%




\end{document}